\RequirePackage{fix-cm}
\documentclass[smallextended]{svjour3}       
\smartqed  
\usepackage{graphicx}
\usepackage{physics}
\begin{document}

\title{The Hoyle Family
}
\subtitle{The search for alpha-condensate states in light nuclei}


\author{R.~Smith$^*$ \and J.~Bishop \and J.~Hirst \and Tz.~Kokalova \and C.~Wheldon
}


\institute{R. Smith \and J. Hirst \at
              Faculty of Science, Technology and Art, Sheffield Hallam University, Sheffield, S1 1WB, UK \\
              R. Smith \at
              The LNS at Avery Point, University of Connecticut, Groton, CT 06340-6097, USA\\
              Tel.: +44 114 225 3430\\
              \email{Robin.Smith@shu.ac.uk}           
           \and
           J. Bishop \at
              Cyclotron Institute, Texas A\&M University, College Station, TX 77843, USA\\
              \email{jackbishop@tamu.edu}
              \and
              Tz. Kokalova \and C. Wheldon \at
              School of Physics and Astronomy, University of Birmingham, Edgbaston, B15 2TT, UK
}

\date{Received: date / Accepted: date}

\maketitle

\begin{abstract}
Our present understanding of the structure of the Hoyle state in $^{12}$C and other near-threshold states in $\alpha$-conjugate nuclei is reviewed in the framework of the $\alpha$-condensate model. The $^{12}$C Hoyle state, in particular, is a candidate for $\alpha$-condensation, due to its large radius and $\alpha$-cluster structure. The predicted features of nuclear $\alpha$-particle condensates are reviewed along with a discussion of their experimental indicators, with a focus on precision break-up measurements. Two experiments are discussed in detail, firstly concerning the break-up of $^{12}$C and then the decays of heavier nuclei. With more theoretical input, and increasingly complex detector setups, precision break-up measurements can, in principle, provide insight into the structures of states in $\alpha$-conjugate nuclei. However, the commonly-held belief that the decay of a condensate state will result in $N$ $\alpha$-particles is challenged. We further conclude that unambiguously characterising excited states built on $\alpha$-condensates is difficult, despite improvements in detector technology. 

\end{abstract}


\section{Introduction}
\label{sec:intro}

The Hoyle state in carbon-12 is considered royalty in the world of nuclear physics. This prestige originates from the crucial role it plays during helium burning, facilitating the production of $^{12}$C through the triple-$\alpha$ process \cite{HoyleStateJenkins}. In order to account for the amount of $^{12}$C and $^{16}$O in the universe, Yorkshire-born astrophysicist Sir Fred Hoyle proposed the existence of a resonance in $^{12}$C, 300 keV above the 3$\alpha$ threshold, required to increase the cross section by seven orders of magnitude \cite{HoyleOriginal}. Under the insistence of Hoyle, the existence of this state was since discovered experimentally \cite{HoyleDisc1,HoyleDisc2}, and hence bears his name. Since then, the Hoyle state has been studied extensively both experimentally and theoretically. The resonance parameters, such as $\Gamma_{\textrm{rad.}}$ and $\Gamma_{\alpha}$, are now mainly well constrained and its role in stellar nucleosynthesis well understood.

Despite this, the \emph{structure} of the Hoyle state is still hotly debated. Owing to its astrophysical role, it is intuitive to think that this particular state in $^{12}$C could, to some level, consist of $\alpha$-particle clusters, whereby the important degrees of freedom are those of $\alpha$-particles, rather than individual nucleons. This is now generally accepted to be the case, however, the exact details of the $\alpha$ interactions and the extent to which their underlying fermion structures play a role is not yet fully understood.

Throughout the history of nuclear physics, the idea of $\alpha$-particle clustering has been present. Predating the discovery of the Hoyle state, in 1938, Hafstad and Teller \cite{HafstadTeller} noted that the ground state binding energies of $N = Z$, $\alpha$-conjugate nuclei, follow a linear relationship with the number of $\alpha$-$\alpha$ bonds, when the proposed $\alpha$-clusters are arranged in crystal-like configurations. The idea of clustering was later extended by Ikeda and colleagues in 1968 \cite{Ikeda}, who suggested that it is necessary for the excitation energy of the nucleus to approach a cluster decay threshold, in order for a structural change into a clustered state. For example, the Hoyle state lies just beyond the 7.27 MeV 3$\alpha$ threshold. These two examples are oversimplifications and we now have experimental evidence of, and theoretical descriptions for, $\alpha$-clustering in both the ground and excited states of nuclei.

The proposed structure of the Hoyle state has had input from other areas of Physics. Since the discovery of atomic Bose-Einstein condensation in 1995 \cite{BECatom}, there has been much speculation about whether similar phenomena may occur in atomic nuclei. Nuclear matter is particularly well-suited for the study of correlation effects in strongly coupled systems of fermions, where the transition from Bardeen-Cooper-Schrieffer (BCS) pairing to Bose-Einstein condensation (BEC) may be investigated. The possibility of $\alpha$-particle condensation in infinite matter has previously been theoretically investigated \cite{InfiniteMatter} and it was found to be possible at low densities (below a fifth of the nuclear saturation density).

The case of finite nuclear systems was approached in a flagship 2001 paper by Tohsaki, Horiuchi, Schuck and R\"{o}pke (THSR) \cite{THSRorig}, who concluded that such a condensate state could exist in light $\alpha$-conjugate nuclei at energies around the $\alpha$-decay threshold. This theoretical approach has played a leading role in the description of near-threshold states in $\alpha$-conjugate nuclei for nearly 20 years. However, an incredibly important issue still remains: the THSR approach reproduces some experimental observables, particularly for $^8$Be and $^{12}$C, though these are not necessarily unique for a condensate.

This review begins by describing the THSR approach and its predictions. Experimental indicators for a condensate are subsequently discussed. Experimental searches for $\alpha$-condensates in a range of nuclei are reviewed, particularly in relation to precision break-up measurements. Future challenges for this theory are finally covered. Open questions include understanding the nature of further, higher energy, excited states in these nuclei, that could correspond to excitations of condensate states.


\section{Alpha cluster models and the THSR wave function}
\label{sec:THSR}

A number of theoretical approaches have been used to study the structure of the $^{12}$C nucleus. State-of-the-art \emph{ab initio} approaches such as Antisymmetrized Molecular Dynamics (AMD) \cite{AMDreview,AMDcarbon}, Fermionic Molecular Dynamics (FMD) \cite{FMD1}, and a calculation on the QCD lattice, utilising Chiral Effective Field Theory \cite{Epelbaum}, have all demonstrated the emergence of $\alpha$-clustering from the nucleon-nucleon interaction. There have also been a number of attempts over the years to understand the structure directly in terms of possible $\alpha$-particle building blocks. The Alpha Cluster Model describes the system in this way, treating $^{12}$C as three quartets, formed from pairs of protons and neutrons in a relative s-wave $-$ $\alpha$-particles. This Alpha Cluster Model was first considered by Margenau \cite{Marenau} and then further developed by Brink \cite{Brink1,Brink2}. The wave functions of each quartet are written as

\begin{eqnarray}
\phi_i(\vec{r},\vec{R}_i) = \sqrt{\frac{1}{b^3\pi^{3/2}}}\textrm{exp} \left [  \frac{-(\vec{r} - \vec{R}_i)}{2b^2} \right],
\label{eq:BrinkWavefunction1}
\end{eqnarray}

\noindent where $b$ is a scaling parameter, which scales with the size of the $\alpha$-particle and $\vec{R}_i$ defines the position of the $i^{\textrm{th}}$ $\alpha$-particle.
Although the $\alpha$-particles themselves are 0$^+$ bosons, the underlying fermion structures require the total wave function of the three $\alpha$-particles to be antisymmetrised as

\begin{eqnarray}
\Phi(\vec{R}_1,\vec{R}_2,\vec{R}_3) = \mathcal{A} \prod_{i=1}^3 \phi_i(\vec{r},\vec{R}_i).
\label{eq:BrinkWavefunction2}
\end{eqnarray}

\noindent For short inter-$\alpha$ distances, the antisymmetrisation breaks the $\alpha$-particle structures, whereas for large $\alpha$-particle separations they retain their bosonic identities.

Possible arrangements of $\alpha$-particles are explored using a variational method. Using a Hamiltonian employing an effective nuclear interaction, the total energy of the system was evaluated as a function of the size and relative positions of the $\alpha$ clusters. Brink found that for $^{12}$C, two structures appear: an equilateral triangle ground state and a 3$\alpha$ linear chain at higher energy (often associated with the Hoyle state). The Algebraic Cluster Model also predicts an equilateral triangle ground state \cite{DanPRL}. However, the prediction of a linear chain is now known to be incorrect for the Hoyle state. Such a spatially extended structure carries a large moment of inertia. Thus, a predicted 2$^+$ rotational excitation of the Hoyle state would appear at a lower energy than the now-measured 10 MeV state \cite{Zimmerman}. This model has also been applied to other light $\alpha$-conjugate systems such as $^{16}$O \cite{Brink16O}. The ground state was calculated to be spherical and excited states were calculated to be strongly $\alpha$-clustered. A series of further calculations of the structure of $^{24}$Mg were also performed by Marsh and Rae \cite{Brink24Mg}.

The alpha cluster model of Brink was refined in 2001 by Tohsaki, Horiuchi, Schuck and R\"{o}pke (THSR) \cite{THSRorig}. They concluded that for states in $^{12}$C with large radii, corresponding to large average $\alpha$-$\alpha$ separations, the $\alpha$-particles may retain their bosonic identities and produce the equivalent of a Bose-Einstein condensate. There is clear evidence indicating that the Hoyle state has an unusually large radius. The form factor for inelastic electron scattering from $^{12}$C has indicated that the volume of the Hoyle state may be up to four times larger than the ground state \cite{HoyleRad1,HoyleRad2,HoyleRad3,HoyleRad4}, depending on the model-dependent analyses employed. Under these conditions, the antisymmetriser in equation \ref{eq:BrinkWavefunction2} will have a weaker effect than on the ground state. In this case, there is a possibility that the larger system could be described, to a good approximation, as a system of three bosons.

The THSR wave function explores this structural possibility and has a similar form to equation \ref{eq:BrinkWavefunction2} beginning as an antisymmetrised product of $\alpha$-particle wave functions.

\begin{eqnarray}
\Phi_{3\alpha} = \mathcal{A} \prod_{i=1}^3 \phi_{\alpha i}(\vec{r_{1i}},\vec{r_{2i}},\vec{r_{3i}},\vec{r_{4i}}).
\label{eq:THSR1}
\end{eqnarray}

\noindent The above construction is for 12 nucleons grouped into quartets described by $\phi_{\alpha i}$. The variables $\vec{r_{1i}}$ etc. denote the coordinates for each nucleon in the $i^{\textrm{th}}$ quartet. The wave functions of each $\alpha$-particle are given as

\begin{eqnarray}
\phi_{\alpha i} (\vec{r_1},\vec{r_2},\vec{r_3},\vec{r_4}) = e^\frac{ -\vec{R} \cdot \vec{R} }{ B^2 }
\exp{ \frac{ -[\vec{r_1} - \vec{r_2}, \vec{r_1} - \vec{r_3} ... ]^2}{ b^2 }},
\label{eq:THSR2}
\end{eqnarray}

\noindent where $\vec{R}$ represents the centre-of-mass coordinate for the quartet. As can be seen, the wave function of each quartet is simply a Gaussian wave packet, spatially modulated by the $\exp{ -\vec{R} \cdot \vec{R} / B^2 }$ factor. The parameter, $b$, still controls the size of each quartet, as in the Brink Alpha Cluster Model, but $B$ is an additional parameter that controls the size of the common Gaussian distribution of the whole nuclear wave function. In the limit that $B \rightarrow \infty$ then the antisymmetrisation $\mathcal{A}$ has no effect and equation \ref{eq:THSR2} simply becomes the product of Gaussian wave packets $-$ a gas of free $\alpha$-particles. Therefore, $B$ is an extra variational parameter and is what makes this treatment of the system so powerful.

\begin{figure*}
\centering
  \includegraphics[width=0.75\textwidth]{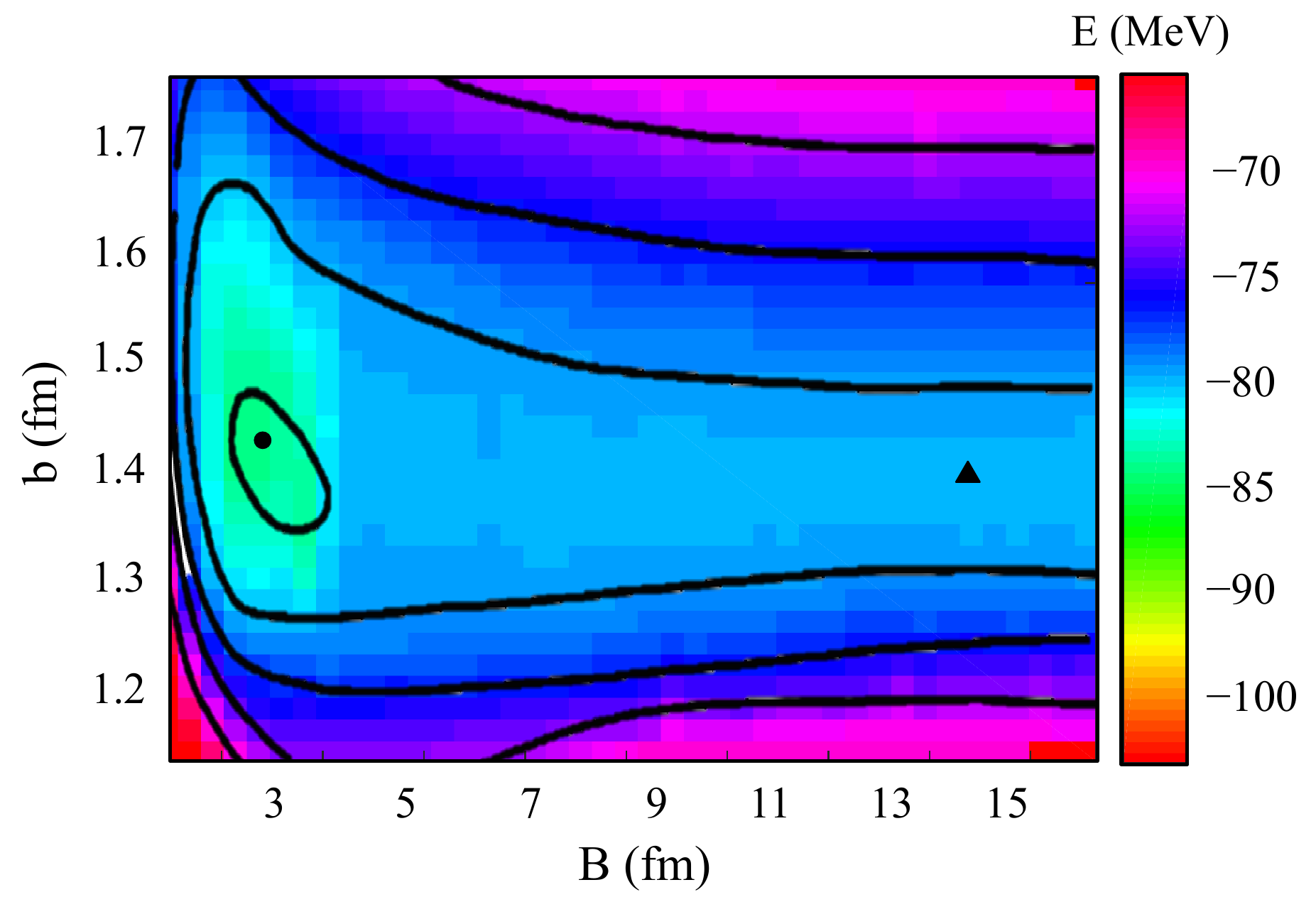}
\caption{Contour map of the energy surface $E_{3\alpha}(B,b)$ for $^{12}$C. The colour map and contour lines denote the binding energies. Data and formulation were obtained from reference \cite{THSRorig}. The circle represents the minimum in the energy surface and the triangle marks a saddle point.}
\label{fig:THSRenergysurface}
\end{figure*}

Possible structures of the nucleus are explored in the same way as the Brink wave function, by performing a variational calculation, this time with both the $b$ and $B$ parameters. The energy surfaces in this two-parameter space can be evaluated as $\expval{\hat{H}}{\Phi_{3\alpha}}$, where the Hamiltonian consists of the kinetic energy, Coulomb energy, and an effective nuclear interaction potential. Various potentials have been used, which give broadly the same features in the energy surfaces. Potentials are chosen that well-reproduce the binding energy and radius of the $\alpha$-particle and the $\alpha$-$\alpha$ scattering phase shifts. The resulting energy surface for $^{12}$C, calculated with the F1 nuclear interaction \cite{F1interaction}, is given in figure \ref{fig:THSRenergysurface}. Equivalent surfaces have been determined for other $\alpha$-conjugate nuclei, $^{8}$Be, $^{16}$O and $^{20}$Ne.

In the case of $^{12}$C, the minimum in the potential energy surface, denoted by the circle in figure \ref{fig:THSRenergysurface}, corresponds closely to the ground state binding energy. The corresponding $b$ and $B$ values at this minimum reproduce the size of the $\alpha$-particle and the compact ground state of $^{12}$C. From the minimum, a ridge is seen extending out towards large values of $B$. The ridge has a saddle point at ($b \approx 1.4$ fm, $B \approx 14$ fm) and has an energy approaching that of the 3$\alpha$ threshold. It is thought that this saddle point, indicated by the triangle in figure \ref{fig:THSRenergysurface}, helps to stabilise a state in $^{12}$C at much larger $B$ values than the ground state. This point could be identified as the Hoyle state given its energy and known large volume compared with the ground state of $^{12}$C. Similar features are seen for other $\alpha$-conjugate nuclei. Therefore, the existence of excited states in these nuclei, with very large volumes compared with the ground states, has been postulated. Given their large volumes, it was proposed that these could correspond to $\alpha$-condensate-type states, with structures well approximated as gases of free $\alpha$-particles.

\begin{figure*}
\centering
  \includegraphics[width=0.75\textwidth]{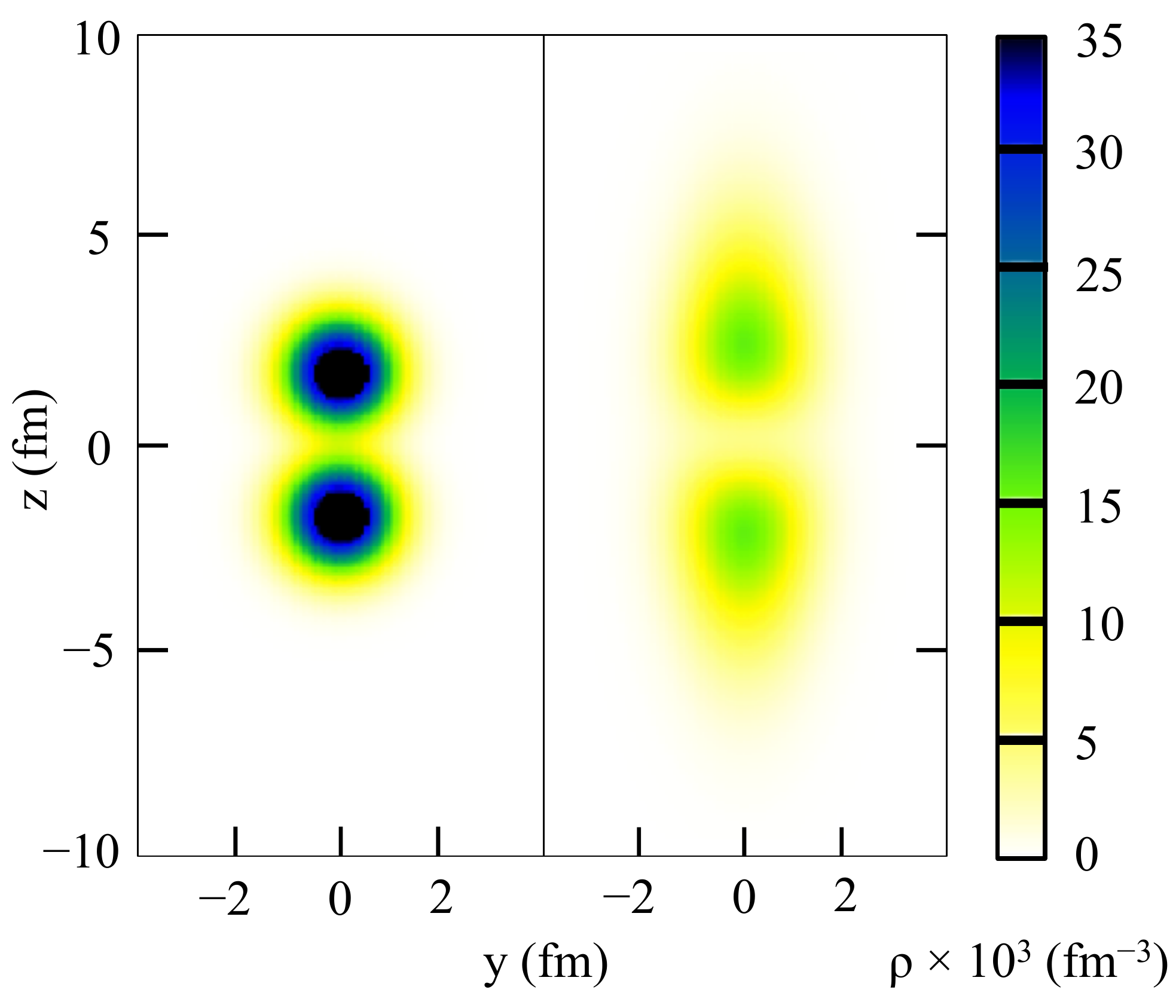}
\caption{Comparison of the intrinsic nucleon densities of the $^{8}$Be ground state calculated using the Brink wave function (left) and THSR wave function (right). Calculations extracted from reference \cite{8BeDensity}.}
\label{fig:8BeDensity}
\end{figure*}

The qualitative similarities and differences between the Brink Alpha Cluster Model and the THSR approach can be seen in figure \ref{fig:8BeDensity}, which shows the intrinsic nucleon densities calculated for the $^{8}$Be ground state. The Alpha Cluster Model effectively places $\alpha$-particles at fixed points in space, giving a 2$\alpha$ dumbbell structure, with an $\alpha$ separation around 4 fm. The THSR model also predicts a similar dumbbell structure. However, apparent stronger repulsion at shorter distances is seen and broad tails appear at larger radii where the Coulomb repulsion is weaker.

One way to explore the possibility of an $\alpha$-condensate-type state arising from the THSR approach is to decompose the calculated Hoyle-state wave function into the single $\alpha$-particle orbitals. Given that the Hoyle state has a large volume, the influence of antisymmetrisation between the $\alpha$-particles should be significantly weakened. In agreement with this picture, the $\alpha$-particle occupation probabilities for the ground and Hoyle state are very different \cite{OccupationNumbers}. There is a 70\% overlap of the Hoyle state THSR wave function with three $\alpha$-particles in the lowest 0$s$-orbital, meaning that the Hoyle state is well approximated by the ideal Bose gas picture. Conversely, the ground state of $^{12}$C is strongly fragmented across $s$, $d$ and $g$ levels, consistent with the shell model. It should be reiterated here that the THSR approach does not advocate that the Hoyle state is a pure $\alpha$-condensate; the fact that there is a 30\% contribution from other orbitals than the 0$s$ indicates that the Pauli Exclusion Principle still plays a significant role. 




\section{Probing the charge distribution}
\label{sec:FormFactor}

One way to experimentally probe the structure of the Hoyle state is to measure the charge distribution  through inelastic electron scattering \cite{HoyleRad1,HoyleRad2,HoyleRad3,HoyleRad4}. As mentioned in section \ref{sec:THSR}, a key prediction of the THSR model is that an $\alpha$-condensate-type state only occurs for volumes much larger than that of the ground state. Therefore, measuring the overlap between the ground and Hoyle states should be a sensitive probe of their structures.

Since the electromagnetic interaction is fully understood, the only unknowns in describing this type of reaction are the nuclear transition charge and current densities. In such experimental measurements, an electron impinges on a $^{12}$C target, populating the Hoyle state. From the cross section and electron momentum distribution, the transition form factor is determined, which provides a clean measure of the overlap between the ground state and the Hoyle state. In electron scattering, the theoretical scattering amplitude due to a point charge is easily evaluated, but must be modified by the form factor for scattering from a finite distribution of charge. The form factor is simply the 3D Fourier transform of the charge distribution, and is given as

\begin{eqnarray}
F(\vec{q}) = \int e^{\vec{q} \cdot \vec{r} / \hbar} \rho(\vec{r}) d^3\vec{r}.
\label{eq:FormFactor}
\end{eqnarray}

\begin{figure*}
\centering
  \includegraphics[width=0.75\textwidth]{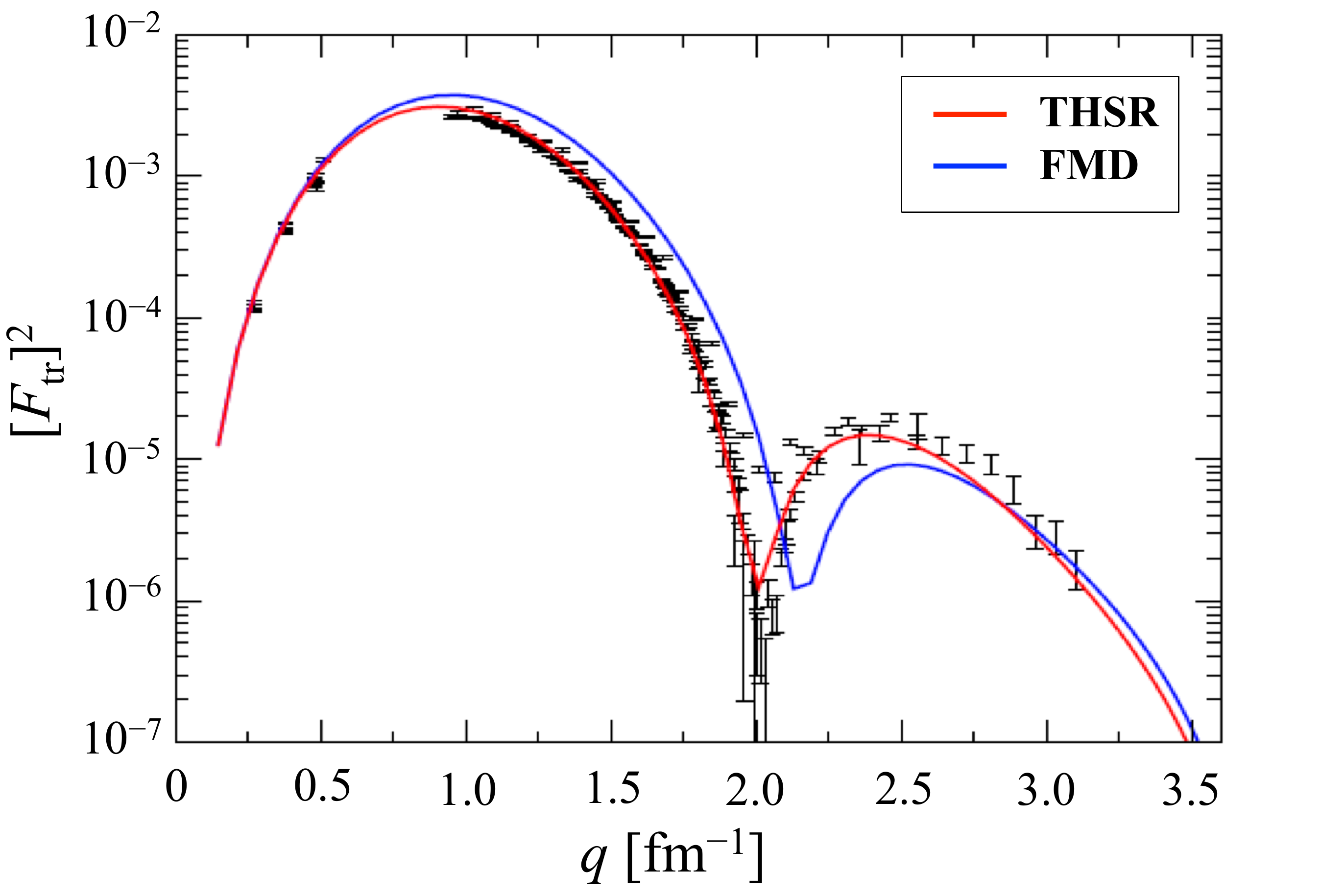}
\caption{Comparison of experimental and calculated inelastic form factors. The solid red line shows the THSR prediction \cite{CondensateFF}, the solid blue line, shifted to slightly higher values of $q$, shows the FMD prediction \cite{FMD1}.}
\label{fig:FormFactor}
\end{figure*}

Analyses of such data are unfortunately not model-independent. Since the E0 monopole interaction depends on the penetration of the incident electron into the nucleus, the plane-wave Born Approximation is fairly inaccurate. Therefore, to determine reduced transition probabilities, as defined in the Born Approximation, the measured inelastic cross sections are converted as

\begin{eqnarray}
\left( \frac{d\sigma}{d\omega} \right )_{\textrm{exp.}} = \left( \frac{d\sigma}{d\omega} \right)_{\textrm{B.A.}}K^2(E_0,q).
\label{eq:FormFactor}
\end{eqnarray}

\noindent where the $K^2(E_0,q)$ factors are determined by comparing the plane-wave Born Approximation with Distorted Wave Born Approximation (DWBA) calculations. Despite small $q$, the influence of higher moments introduces an intractable systematic uncertainty in the measured form factors. Experimental measurements of the derived form factors for transitions from the ground state to the Hoyle state are shown in figure \ref{fig:FormFactor}. Five sets of inelastic electron scattering data, from four different laboratories, were globally analysed in reference \cite{ElectronInelastic}, covering a range in $q$ from 0.27 to 3.04 fm$^{-1}$.

To interpret the experimental measurements, theoretical models are required that can describe both the ground and Hoyle states of $^{12}$C, since their overlap must be evaluated. Both the THSR \cite{CondensateFF} and FMD models \cite{FMD1} calculate the ground and Hoyle state wave functions, each indicating that the Hoyle state has a radius larger than that of the ground state by a substantial factor. Green's function Monte Carlo (GFMC) calculations \cite{GFMCFF} also well reproduce the inelastic form factor. The THSR and FMD models are compared with experimental data in figure \ref{fig:FormFactor}. The GFMC calculations \cite{GFMCFF} could not be plotted for comparison on the same scale. The THSR and GFMC fit the data extremely well, whereas the FMD calculations do not. However, the FMD and GFMC calculations underbind the Hoyle state by 2$-$2.5 MeV relative to the ground state. On the other hand, the THSR wave function calculates an excitation energy much closer to the experimental value of 7.65 MeV. In reference \cite{Zhou2020}, using Volkov No. 2 forces, an excitation energy of 7.73 MeV was obtained, 250 keV above the calculated 3$\alpha$ threshold.

Based on the excellent fit to the experimental data, it appears that the THSR model well describes the structures of the ground and Hoyle states of $^{12}$C. Remarkably, this excellent fit is obtained with no tuneable parameters. The close agreement with GFMC and FMD approaches demonstrates that the approximate $\alpha$-condensate nature of the Hoyle state, predicted by the THSR model, also arises naturally in these other \emph{ab initio} approaches. It should be mentioned here that the experimental searches for a Hoyle state equivalent in $^{16}$O have never utilised the form factor as a way to confirm the nature of this state.


\section{Precision break-up measurements}
\label{sec:Breakup}

The form factor for inelastic electron scattering is a clear, albeit not model-independent, way to measure the overlap between the Hoyle state and the ground state of $^{12}$C. The THSR $\alpha$-condensate model describes this experimental observable very well. However, only so much weight can be given to a single observable.

Precision break-up measurements of the Hoyle state into three $\alpha$-particles should provide a complimentary way to determine the nature of this state. In 2006, Tz. Kokalova and colleagues \cite{KokalovaCondensate} concluded that the branching ratios for various decay channels of a nuclear state could provide direct signatures for $\alpha$-condensation. The decay of a possible $\alpha$-condensed state will consist of a variety of decay modes. For example, the $0_{6}^{+}$ resonance in $^{16}$O at 15.1 MeV, which has been proposed as a Hoyle state analogue, can decay through the following channels:

\begin{eqnarray}
\label{eq:Breakup16O1}
^{16}\textrm{O}_{0_6^+} &\rightarrow& ^{12}\textrm{C}_{0_1^+} + \alpha\\
			\label{eq:Breakup16O2}
                         &\rightarrow& ^{12}\textrm{C}_{2_1^+} + \alpha\\
                         \label{eq:Breakup16O3}
                         &\rightarrow& ^{12}\textrm{C}_{0_2^+} + \alpha\\
                         \label{eq:Breakup16O4}
                         &\rightarrow& ^{8}\textrm{Be}_{0_1^+} +^{8}\textrm{Be}_{0_1^+}\\
                         \label{eq:Breakup16O5}
                         &\rightarrow& ^{8}\textrm{Be}_{0_2^+} +^{8}\textrm{Be}_{0_2^+}\\
                         \label{eq:Breakup16O6}
                         &\rightarrow& ^{8}\textrm{Be}_{0_1^+} +^{8}\textrm{Be}_{0_2^+}\\
                         \label{eq:Breakup16O7}
                         &\rightarrow& \alpha + \alpha + \alpha + \alpha.
\end{eqnarray}

\noindent For the Hoyle state, the only open channels are:

\begin{eqnarray}
\label{eq:Breakup12C1}
^{12}\textrm{C}_{0_2^+}  &\rightarrow& ^{8}\textrm{Be}_{0_1^+} + \alpha\\
\label{eq:Breakup12C2}
                         &\rightarrow& ^{8}\textrm{Be}_{2_1^+} + \alpha\\
                         \label{eq:Breakup12C3}
                         &\rightarrow& \alpha + \alpha + \alpha.
\end{eqnarray}

If the decaying nuclear state is an $\alpha$-condensed state, all of the $\alpha$-clusters occupy the same 0$s$ orbit. This means that any partitioning of the nucleus into subsystems, which are also $\alpha$-condensed states, is possible, and should be equally probable. Therefore, in the case of $^{16}$O, channels (\ref{eq:Breakup16O3}), (\ref{eq:Breakup16O4}) and (\ref{eq:Breakup16O7}) should be equally probable, since these decays proceed through proposed $\alpha$-condensed states in $^{12}$C and $^8$Be. This means that the experimentally measured channel widths/branching ratios will be determined only by the phase space available for each decay and the penetrability through the Coulomb barrier. By the same argument, in the case of $^{12}$C, the sequential decay (\ref{eq:Breakup12C1}) and direct 3$\alpha$ decay (\ref{eq:Breakup12C3}) should be equally probable, since the $^{8}\textrm{Be}_{0_1^+}$ is thought to be an $\alpha$-condensate. Therefore, their corresponding relative decay widths should be entirely calculable from phase space and Coulomb barrier penetrabilities.

\subsection{Carbon-12}
\label{sec:carbon12}

Much experimental effort has been devoted to measuring the 3$\alpha$ direct decay width of the Hoyle state in recent years \cite{KirsebomHoyle,ItohHoyle,SmithHoyle,NapoliHoyle,RanaHoyle}. The current section focuses on the data of reference \cite{SmithHoyle}, first published as a letter in 2017, followed by several articles for a non-specialist audience \cite{KirsebomViewpoint,PhysicsWorld}. A major issue in determining the 3$\alpha$ direct decay width is that the phase space for direct decay (\ref{eq:Breakup12C3}) is so much smaller than for the sequential decay (\ref{eq:Breakup12C1}). The phase spaces are calculable using the Fermi breakup model \cite{PhaseSpace} and the direct decay is suppressed by a factor of 10$^3$ relative to the sequential decay. This means that measuring the direct decay requires very high statistics data. At present, an upper limit of 0.0019\% has been placed on the 3$\alpha$ direct decay branching ratio, utilising around 2 $\times$ 10$^4$ Hoyle state decay events \cite{RanaHoyle}.

In such experiments, a beam of particles, such as $\alpha$-particles, inelastically scatter from a $^{12}$C target, populating the Hoyle state in the recoiling carbon nucleus. Transfer reactions have also been used \cite{KirsebomHoyle,NapoliHoyle}. The excited $^{12}$C then decays into three $\alpha$-particles, which hit position-sensitive silicon strip detectors. For this type of experiment, the sequential and direct decay channels are separated by examining the relative energies of the three $\alpha$-particles in the final state. To further complicate the problem, since the Hoyle state is only 380 keV above the 3$\alpha$ threshold, after the decay, these three $\alpha$-particles have very similar energies. Silicon charged-particle detectors typically have an absolute energy resolution of 30$-$50 keV, meaning that differentiating between the three $\alpha$-particles using such detectors can be difficult. An alternative approach is to measure the decay of the Hoyle state in a Time Projection Chamber (TPC) such as those in references \cite{JINST,TexAT,JackCocoyoc}. In these cases, the relative angles between the three $\alpha$-particles could be used to differentiate the two decay channels. Experimental work using this approach is ongoing.

\begin{figure*}
\centering
  \includegraphics[width=0.9\textwidth]{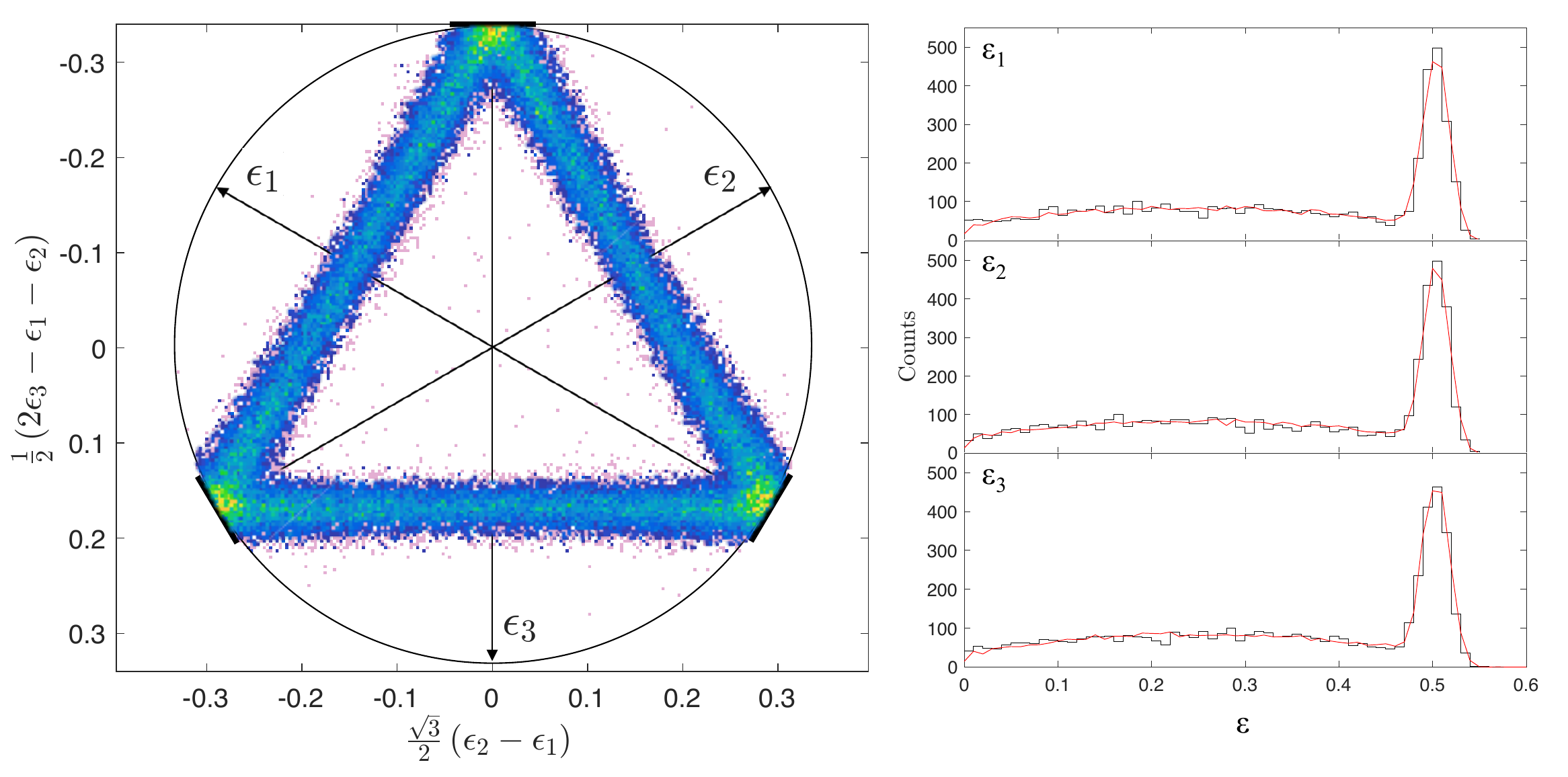}
\caption{Hoyle state decay data from references \cite{SmithHoyle} and \cite{SmithThesis}. Right panel: Histograms of $\alpha$-particle fractional energies. Left panel: $\alpha$-particle fractional energies plotted as a Dalitz plot.
}
\label{fig:DalitzPlots}
\end{figure*}

In the typical analysis approach, the relative energies of the three $\alpha$-particles are examined using a Dalitz plot. In the centre-of-mass of the decaying $^{12}$C, the \emph{fractional energies} of the $\alpha$-particles, $\epsilon_{\alpha_i} = E_{\alpha_i}/E_{tot}$, should all sum to unity. This restriction on the sum of the three fractional energies allows them to be plotted on a two-dimensional symmetric Dalitz plot \cite{DalitzOrig}. The construction of a symmetric Dalitz plot is described in detail in reference \cite{NSD2015Proc}. For the decay of the Hoyle state, the decay kinematics dictate that the first emitted $\alpha$-particle carries away a fixed amount of energy (around 1/2 of the total available) and the remaining energy is shared between the other two $\alpha$-particles. This means that sequential decays appear as a \emph{triangle} on the Dalitz plot. A subset of the experimental data from reference \cite{SmithHoyle} are shown in figure \ref{fig:DalitzPlots}. The right panel shows three 1D histograms of $\alpha$-particle fractional energies and the left panel shows the same data plotted as a Dalitz plot.

Higher-dimensional Dalitz plots are also possible, in order to examine the $N\alpha$ decays of $^{16}$O and heavier nuclei, although such analyses have not yet been performed. However, a three-dimensional Dalitz plot has previously been used in atomic physics to understand 4-body atomic break-up processes \cite{AtomDalitz}.

As can be seen in the left panel of figure \ref{fig:DalitzPlots}, the vast majority of data lie on a triangle, indicating a dominant sequential decay, as expected from the relative phase spaces. A small number of counts beyond this triangle can be seen, which could correspond to direct decays. Other alternatives are experimental backgrounds, such as event mixing or mis-assigning hit positions of the $\alpha$-particles on the detectors. To explore the relative amounts of sequential and direct decay, high statistics Monte-Carlo simulations of the experiment were performed, which included background effects. Each decay type $-$ sequential and direct $-$ were simulated, and the resulting Dalitz plot distributions were compared with the experimental data, as a function of the direct decay branching ratio.

\begin{figure*}
\centering
  \includegraphics[width=0.9\textwidth]{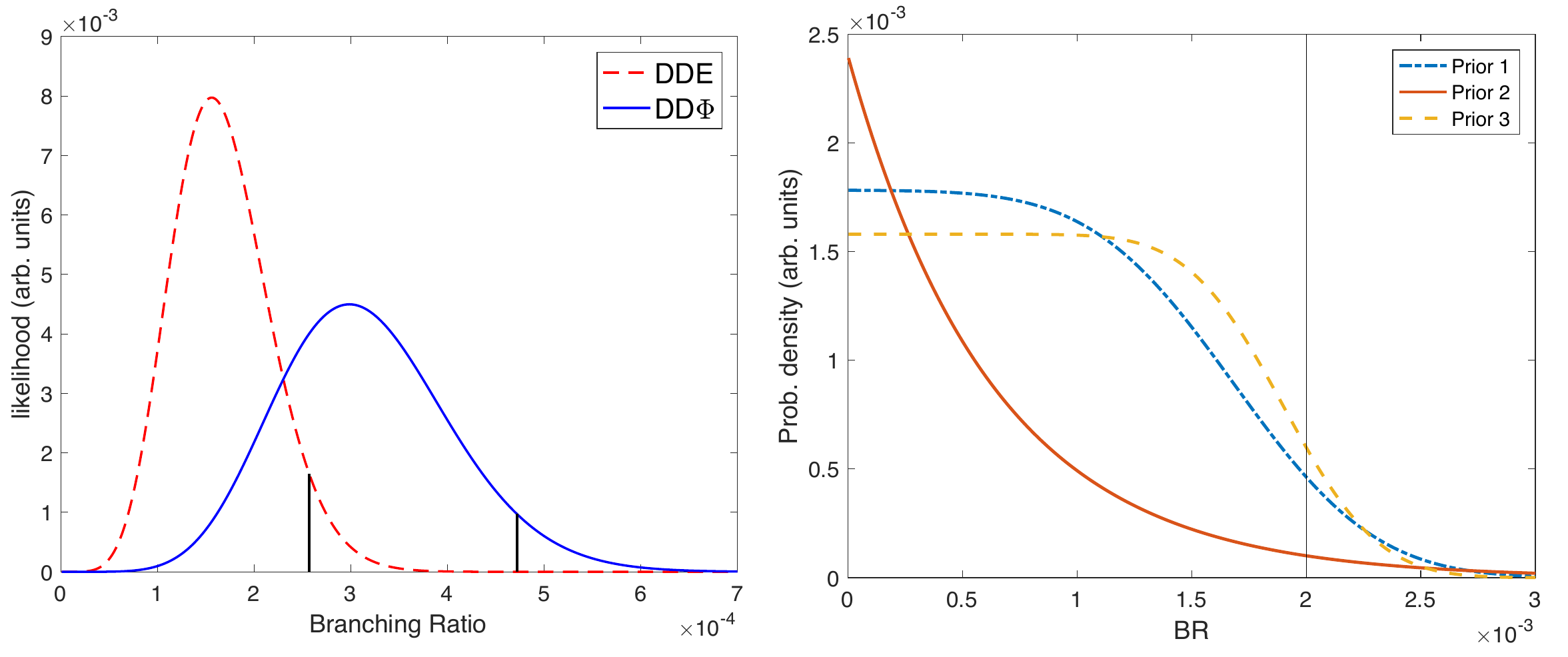}
\caption{Left panel: Likelihood distribution as a function of the direct decay branching ratio. The solid blue line shows the likelihood distribution for a direct decay with equal probabilities to the entire phase space. The red dashed line shows the distribution for a direct decay to equal $\alpha$-particle energies. Right panel: Various prior distributions used in the Bayesian analysis.}
\label{fig:StatisticalAnalysis}
\end{figure*}

The extracted branching ratio from this analysis is clearly sensitive to the exact direct decay model that was simulated, and this will be discussed more later. However, the standard approach is to model an equal probabilities decay to anywhere in the available phase space. This decay type is typically denoted as DD$\Phi$. Such a decay corresponds to a flat distribution of points inside the kinematically allowed circular region of the Dalitz plot (indicated in figure \ref{fig:DalitzPlots}). The theoretical distributions, simulated through Monte-Carlo, were fit to the data using a frequentist approach and further details are given in references \cite{SmithHoyle} and \cite{SmithThesis}. With a 3$\alpha$ direct decay branching ratio of 0\%, a $\chi^2$/dof value of 1.08 was obtained, close to the 50\% confidence level (C.L.). The branching ratio was increased and the $\chi^2$ value moved beyond the 95\% (2$\sigma$) C.L. at a value of 0.0470\%. The upper limit for the direct decay branching ratio was thus placed at 0.0470\% (4.70 $\times$ 10$^{-4}$). This information is captured by the blue likelihood distribution in the left panel of figure \ref{fig:StatisticalAnalysis}. The vertical black line indicates the 2$\sigma$ C.L.

A complementary Bayesian approach was also used to extract an upper limit for the branching ratio. The Bayesian approach rightly asserts that we should not treat the direct 3$\alpha$ decay branching ratio of the Hoyle state as a completely unknown parameter, since a measurement previous to the experiment in question had set an upper limit for direct decay of 0.2\%, at the 95\% C.L. \cite{ItohHoyle}. Therefore, we know with 95\% confidence that the branching ratio is less that 0.2\%. The idea behind the Bayesian analysis was to combine the previous results with the latest experimental measurements in order to better constrain the direct decay branching ratio. This is achieved by defining a prior likelihood distribution for the branching ratio that satisfies the statistical analysis of reference \cite{ItohHoyle}. Specifying the prior distribution is a controversial topic, due to the obvious influence it has on the result. However, in this work, the result was seen to be fairly insensitive to the choice of prior distribution. The prior distributions used in this analysis are shown in the right panel of figure \ref{fig:StatisticalAnalysis}. The Bayesian analysis is built on Baye's Theorem, which states, in the context of this work

\begin{eqnarray}
P(BR|X) = \frac{P(X|BR) P(BR)}{P(X)}.
\label{eq:BayesTheorem}
\end{eqnarray}

\noindent Here, the desired quantity, $P(BR|X)$, represents the probability of a particular branching ratio, $BR$, given the data, $X$. The $P(X|BR)$ represents the probability of obtaining data, $X$, given a certain value of the $BR$, which may be identified as the standard likelihood distribution, shown in the left panel of figure \ref{fig:StatisticalAnalysis}. The $P(BR)$ is the aforementioned prior likelihood distribution for the branching ratio. The $P(X)$ factor is adjusted such that the distribution $P(BR|X)$ is normalised to unity. Utilising this method, a slightly lower branching ratio of 0.0465\% (4.65 $\times$ 10$^{-4}$), was obtained. We advocate that future experimental analyses utilise a similar Bayesian approach.

As previously mentioned, the result is highly sensitive to the simulated direct decay model. An equal probabilities decay to the phase space is typically utilised, but other models do exist. One is the DDE direct decay model, where the $\alpha$-particles are emitted with equal energies. This corresponds to the point at the centre of the Dalitz plot. We have previously argued that this cannot always be the case \cite{SmithThesis}; due to the finite size of the decaying Hoyle state, Heisenberg's position-momentum uncertainty principle will smear the kinetic energies of the emitted $\alpha$-particles. Another direct decay type is the DDL model. Time-dependent Hartree-Fock calculations \cite{TDHFStevenson} have demonstrated that a linear chain state of three $\alpha$-particles in $^{12}$C can be produced through the triple-$\alpha$ process. However, a stable configuration only occurs if the third $\alpha$ strikes the $^8$Be with a small impact parameter along the direction of $^8$Be deformation. It is natural then to conclude that during the decay of the Hoyle state, if it is indeed a linear chain of $\alpha$-particles, that they would be emitted from the nucleus in a collinear way. This type of decay corresponds to points on the outer edge of the Dalitz plot. A final model, developed in references \cite{SmithHoyle} and \cite{CocoyocProceedings}, is called DDP$^2$ (Direct Decay Phase space + Penetrability). This model accounts for the changing 3$\alpha$ decay penetrability depending on the relative energies and directions of the $\alpha$-particles as they tunnel from the nucleus. Its similarity in results to an $R$-matrix model of the direct decay have previously been noted \cite{Refsgaard2018}. In this model, it is calculated that the Coulomb barrier for an equal energies DDE decay is significantly lower than for a collinear DDL decay. Therefore, the phase space distribution of $\alpha$-particle energies should be non-uniform and peaked towards the centre of the Dalitz plot. Upper limits on the direct decay BR for each model are summarised in table \ref{tab:BRTable}.

\begin{table}
\centering
\begin{tabular}{ c | c c c c}
          & 95\% C.L. & 99.5\% C.L. & 95\% C.L. & 99.5\% C.L.\\
          &  &  & (Bayesian) & (Bayesian)\\
         \hline
         DD$\Phi$ & $4.7 \times 10^{-4}$ & $5.8 \times 10^{-4}$ & $(4.65 \pm 0.05) \times 10^{-4}$ & $(5.67 \pm 0.1) \times 10^{-4}$\\
         DDE/DDP$^2$ & $2.57 \times 10^{-4}$ & $3.2 \times 10^{-4}$ & $-$ & $-$\\
         DDL & $3.8 \times 10^{-5}$ & $6.4 \times 10^{-5}$ & $-$ & $-$\\
\end{tabular}
\caption{The values of branching ratio upper limits for each of the direct decay mechanisms described in the text. The quoted systematic uncertainties are due to the choice of prior distribution in the Bayesian analysis.}
\label{tab:BRTable}
\end{table}

Surprisingly, there are very few theoretical predictions of the direct decay branching that can be compared with the experimental data. In 2014, Ishikawa utilised a full three-body quantum mechanical formulation to study the decay of the Hoyle state \cite{IshikawaHoyle}. In that work, the Hoyle state was treated as a system of three bosonic $\alpha$-particles, thus reflecting an $\alpha$-condensate-type structure. Ishikawa concluded that the direct decay contributes at a level lower than 0.1\%. The latest experimental measurements \cite{RanaHoyle} reject a direct decay contribution $>$~0.019\%, which is an order of magnitude lower than this prediction.

\begin{figure*}
\centering
  \includegraphics[width=0.9\textwidth]{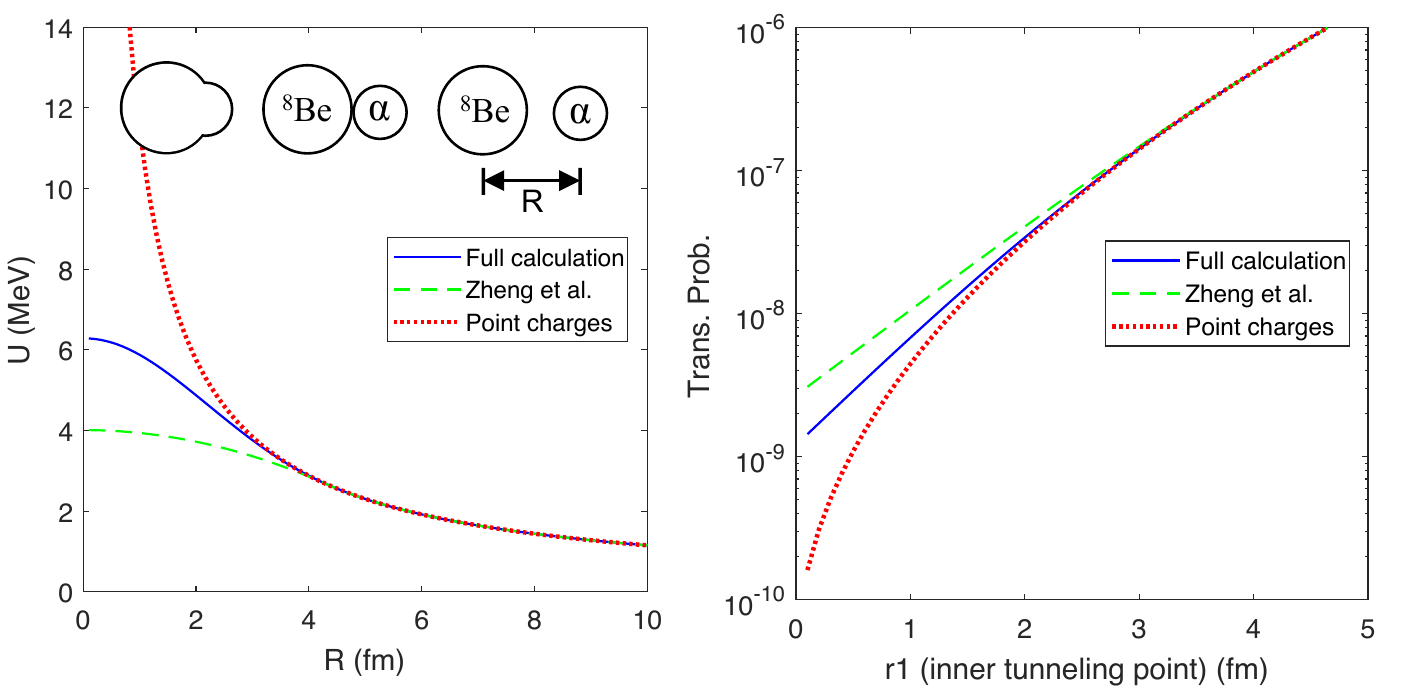}
\caption{Left panel: Comparison between Coulomb potentials commonly utilised in tunnelling calculations. The dotted line shows the Coulomb interaction between point charges, the dashed line shows the potential quoted in references \cite{ZhengHoyle,BarrierParam}, and the solid line shows the full calculation. Right panel: The transmission probability as a function of $r_1$, the inner separation at which the $^8$Be + $\alpha$ tunnel from. Calculations by J. Hirst.}
\label{fig:CoulombBarriers}
\end{figure*}

A simple approach to theoretically determining the BR is to evaluate the relative sequential and direct decay widths using tunnelling calculations. References \cite{SmithHoyle} and \cite{CocoyocProceedings} present WKB calculations for the 2-body and 3-body decays, which calculate the BR to be around 0.06\%. This is higher than the current experimental upper limit. In this model, the Coulomb barrier is treated as that of point charges that tunnel out from the channel radius. This method utilises the PeTA WKB code \cite{PeTA}, which Monte-Carlo samples the allowed phase space to calculate an average Coulomb penetration factor.

In a similar approach, Zheng et al. \cite{ZhengHoyle} performed WKB calculations of tunnelling through a Coulomb potential. However, they used the Gamow prescription, which neglected the nuclear potential. Inclusion of a nuclear interaction would modify the results, as this strongly influences the barrier shape. A branching ratio of 0.0036\% was calculated; considerably below current experimental limits. However, in their paper, they only consider DDE-type decays because ``\emph{We expect a change less than a factor of 2 [by] adding more configurations}". In contrast to this, the 3$\alpha$ phase space distributions calculated in references \cite{Refsgaard2018} and \cite{CocoyocProceedings} demonstrate a large dependence of the barrier transmission probability on the relative energies/orientations of the three $\alpha$-particles. Furthermore, the Coulomb interaction chosen by Zheng et al. was modified to reflect the potential energy of two overlapping, uniformly charged spheres, parameterised for a 2-body decay as

\begin{eqnarray}
\label{eq:CoulombBarrierIn}
U(R) &=& \frac{Z_a Z_b e^2}{2(R_a + R_b)} \left ( 3 - \frac{R^2}{(R_a + R_b)^2} \right ) \hspace{0.2cm} (R \leq R_a + R_b)\\
\label{eq:CoulombBarrierOut}
&=& \frac{Z_a Z_b e^2}{R} \hspace{3.85cm} (R > R_a + R_b).
\end{eqnarray}

\noindent where $Z_i$ and $R_i$ are the charges and radii of each fragment, and $R$ is the separation between their centres. This is a commonly used potential and can also be found quoted in reference \cite{BarrierParam}. However, we demonstrate that this is incorrect. The left panel of figure \ref{fig:CoulombBarriers} shows this potential as a green dashed line, for the decay of $^{12}$C into $^8$Be + $\alpha$. The solid blue line shows the correct potential for the system, determined computationally by integrating over the charge distributions of two overlapping spheres. The difference between the two models is small at the channel radius, but becomes more significant as the two objects overlap. Due to this difference, the barrier transmission probabilities, calculated with WKB, vary significantly as shown in the right panel of figure \ref{fig:CoulombBarriers}. The difference is largest as the inner tunnelling point tends to zero (Gamow limit). We therefore encourage the calculations of \cite{ZhengHoyle} to be performed with the correct potential, although this will probably give a small correction to the result.

In summary, experiments to measure a 3$\alpha$ direct decay width of the $^{12}$C Hoyle state are reaching the limits of what is feasible with current technologies. At present, the only way to improve the situation is by running longer experiments and gaining higher statistics. Experiments utilising TPC detectors, rather than silicons, are underway, but the same problem remains. Additionally, in these systems, scattering of the very low-energy $\alpha$-particles in the gas is an issue. At the same time, theoretical descriptions of the break-up process require further work. We have highlighted issues with the simplistic tunnelling models currently used to evaluate the approximate branching ratio. The THSR and FMD models accurately predict some experimental observables. Can they predict the direct decay branching ratio?

\subsection{Oxygen-16}
\label{sec:Oxygen16}

As stated earlier, theoretical investigations of the Hoyle state in $^{12}$C have established that it is well approximated as a dilute gas-like state of three $\alpha$-particles. Subsequently, there is no reason why there should not exist a whole family of Hoyle analogue states in heavier nuclei. Thankfully, much like the British royal family, such states have a rather small gene pool; they are restricted to $\alpha$-conjugate nuclei such as $^{16}$O, $^{20}$Ne, $^{24}$Mg etc. and have been predicted to have a maximum mass corresponding to $^{40}$Ca \cite{MaxMassCondensate}.

For heavier N$\alpha$ systems, one can again look to break-up measurements as signatures of $\alpha$-condensation. Of particular interest is the 15.1 MeV $0_{6}^{+}$ state in $^{16}\mathrm{O}$, which has previously been measured in the $\alpha_0$ and $\alpha_1$ channels \cite{LiiThemba}. However, the contribution from other states around this energy region is still not well understood and this state has not yet been conclusively demonstrated to correspond to a clustered state. An ideal demonstration of the clustered nature, and in particular of the $\alpha$-condensate nature would be to observe an enhanced sequential $\alpha$-particle emission from one $\alpha$-condensate state to another. To do this, a high-energy compound nucleus reaction $^{12}\mathrm{C}(^{16}\mathrm{O},{^{28}\mathrm{Si}^{\star}})$ was employed at beam energies of 160, 280 and 400 MeV, to populate a wide range of states in $^{8}\mathrm{Be}$ to $^{28}\mathrm{Si}$ \cite{PhysRevC.100.034320}. By looking at the complete decay to a 7 $\alpha$-particle final state, a direct search for N$\alpha$ condensate states was performed by examining their complete dissociation into an N$\alpha$-particle final state.

\begin{figure}
\centerline{\includegraphics[width=0.49\textwidth]{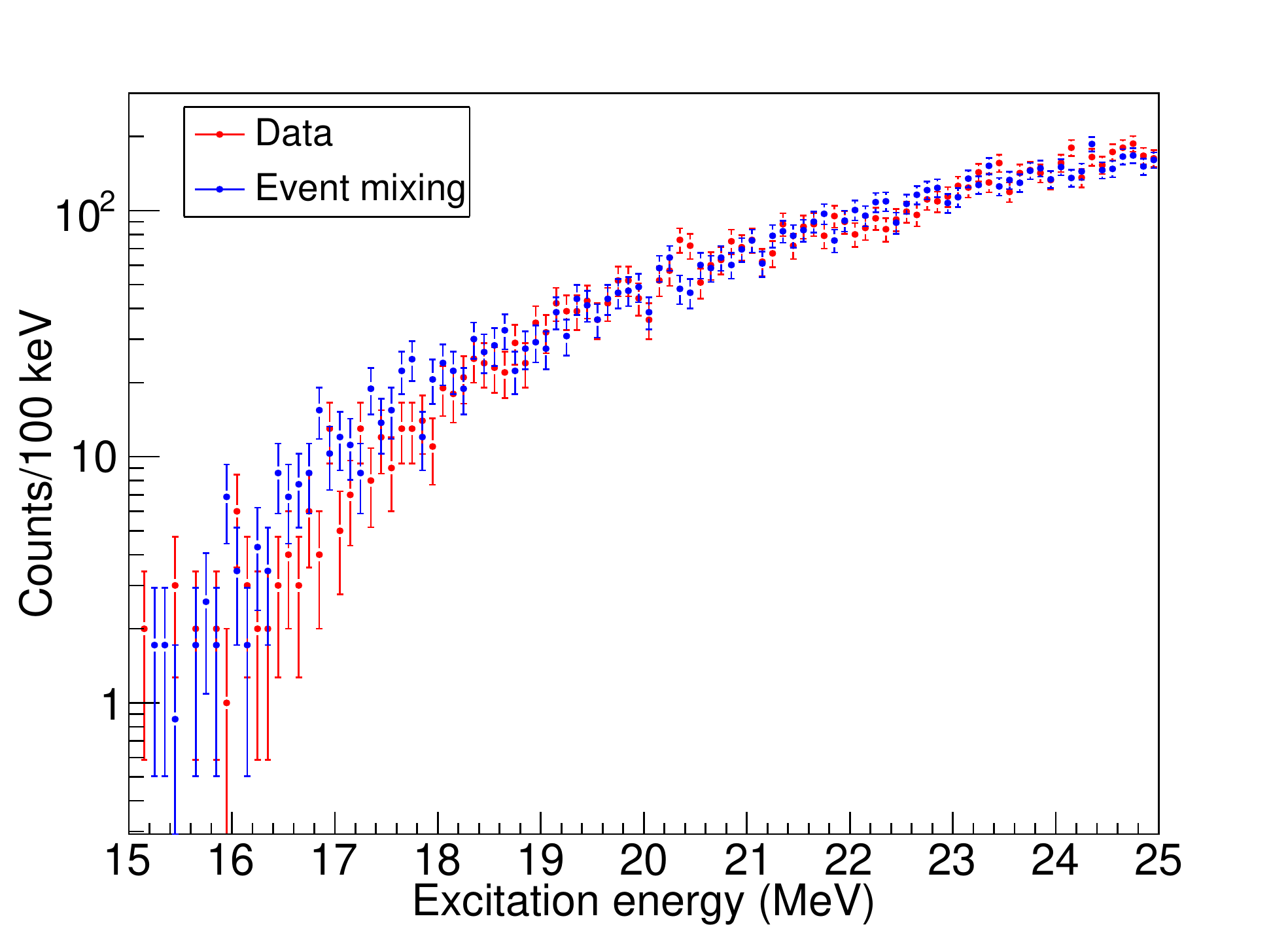}\includegraphics[width=0.49\textwidth]{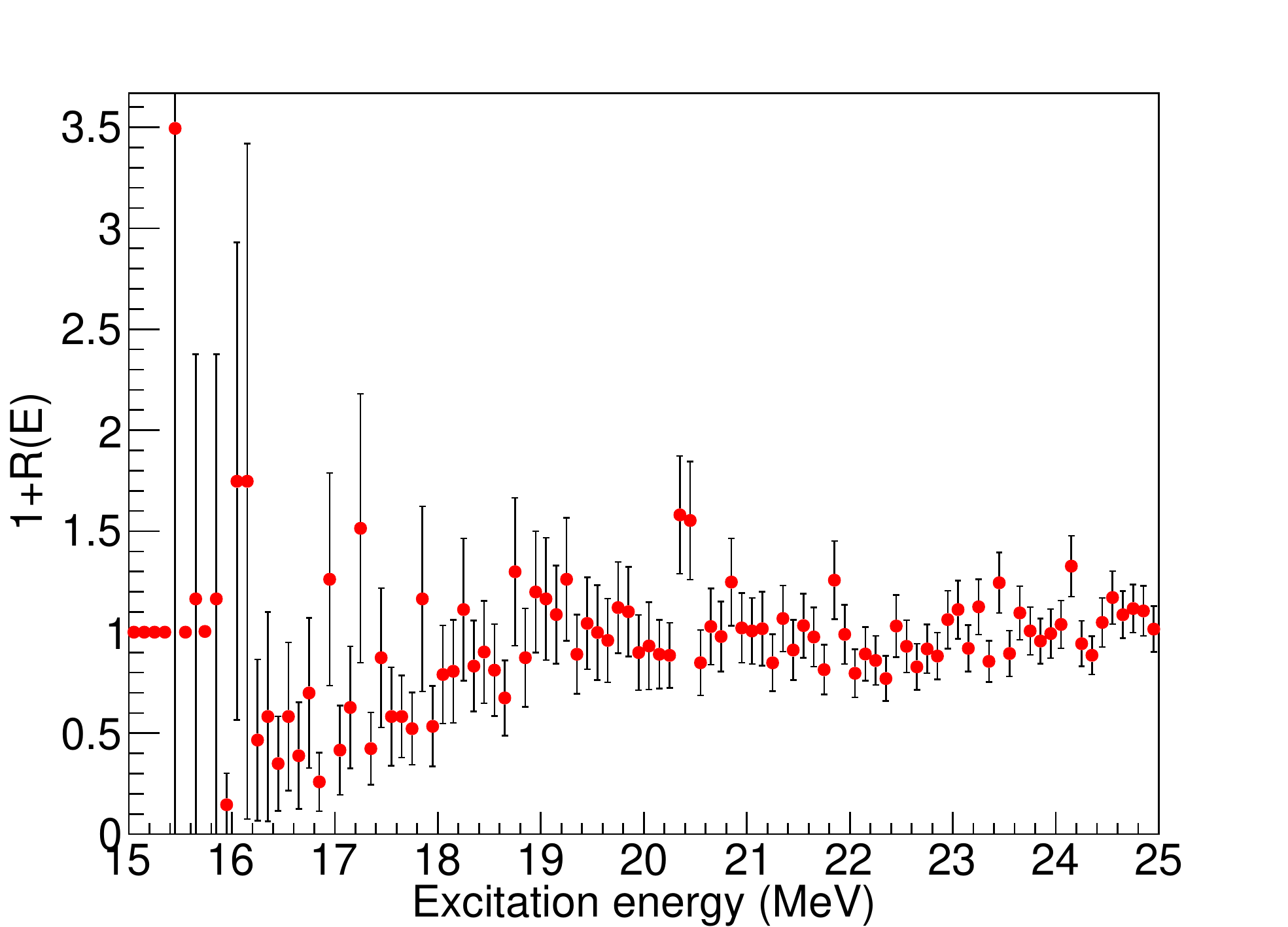}}

\caption{Left panel: $^{16}\mathrm{O}$ excitation energy reconstructed from 4 $\alpha$-particles in the $^{12}\mathrm{C}(^{16}\mathrm{O},4\alpha)$ channel, with a beam energy of 160 MeV. The data (red) are compared to the mixed events (blue) \cite{Mixing}. The mixed events describe the data down to 15 MeV very well. The small number of counts observed around 15 MeV can therefore be assigned to uncorrelated $\alpha$-particles. Right panel: Correlation function of the plot on the left where the ratio of the data to the event mixed data are taken. Any resonances would deviate strongly from unity. While a large correlation value can be seen at small excitation energy (15.5 MeV), the errors demonstrate this is most likely a statistical fluctuation. As such there is no evidence for a state here in the 4$\alpha$ channel.\label{fig:mixing}}
\end{figure}

It was demonstrated that due to the effect of the Coulomb barrier in the decay of $^{16}\mathrm{O}^{\star} \rightarrow 4\alpha$, this decay mode is suppressed up until $\sim$ 18 MeV (in agreement with previous experiments \cite{PhysRevC.51.1682,PhysRevC.70.064311,PhysRevC.88.064309,PhysRevC.94.034313,Bruno_2019}). This means that even with the reduced Coulomb barrier from a dilute $0_{6}^{+}$ state, the decay of this state into 4$\alpha$ is heavily suppressed. As such, this characteristic decay mode cannot be identified \cite{Funaki}. There was no evidence of a state at 15.1 MeV in the 4$\alpha$ channel (see figure~\ref{fig:mixing}), in agreement with some previous results \cite{doi:10.1142/S021773231000068X} and disagreeing with others \cite{Marina}. In the previous study that claimed to find the state \cite{Marina}, no evidence of the effect of the Coulomb barrier was seen in the excitation function, which suggests that mismatched $\alpha$-particles, poor energy resolution and low statistics may be responsible for the observed yield. Additionally, a second measurement at lower energy did not see a peak in the same location.

In the $^{12}\mathrm{C}(^{16}\mathrm{O},{^{28}\mathrm{Si}^{\star}})$ study \cite{PhysRevC.100.034320}, to overcome the limitations of the 4$\alpha$ penetrability, populating this state in the $^{12}\mathrm{C}(^{16}\mathrm{O},{^{12}\mathrm{C}(0_{2}^{+}}))^{16}\mathrm{O}^{\star}$ reaction was attempted, by reconstructing the $^{16}\mathrm{O}^{\star}$ from measuring the $^{12}\mathrm{C}(0_{2}^{+})$. From the compound nucleus, if one decay product ($^{12}\mathrm{C}(0_{2}^{+}$)) is produced which is heavily clustered, one would expect the other decay product to also be preferentially populated by heavily-clustered states. There was no evidence of the population of a state around 15 MeV using this technique.

As discussed above in section \ref{sec:carbon12}, one may also identify an $\alpha$-condensate state by verifying the equivalency of all the $\alpha$-condensate decay modes. To test for evidence of $\alpha$-condensates at higher energies in $^{28}\mathrm{Si}$, the Fermi breakup model was used to calculate the expected yields of 8 different partitions to $\alpha$-condensate states. While this model ignores the penetrability, which has a small effect due to the large relative energy above the barrier, it was shown that the seen experimental yields were not commensurate with an $\alpha$-condensate. Additionally, the Fermi breakup results were used in conjunction with an extended Hauser Feshbach calculation to investigate the role of sequential decay against multi-particle decay. Previous experiments \cite{Akimune_2013} have claimed that a larger-than-expected $\alpha$-multiplicity from the compound nucleus is indicative of $\alpha$-condensation in much heavier systems ($^{56}\mathrm{Ni}$).

It was demonstrated that while the predicted $\alpha$-particle multiplicities from the Hauser Feshbach calculation cannot explain the experimentally observed yields at the three different energies, the Fermi breakup model calculations also incorrectly predicted a peak $\alpha$-particle multiplicity of 4-6 as the beam energy increased. The results of this work therefore do not see any signatures of $\alpha$-condensates and also highlight the importance of understanding the reaction mechanisms involved. The Coulomb barrier suppression is very restrictive for the nuclei studied. Moving to heavier systems where such an $\alpha$-condensate is lightly bound (e.g. $^{40}\mathrm{Ca}$), observing the complete dissociation in a ``Coulomb explosion'' may present the clearest observable of $\alpha$-condensation in heavy systems \cite{MaxMassCondensate}.


\section{Conclusions and outlook}
\label{sec:Conclusions}

Theoretical investigations have established that the Hoyle state is well approximated as a dilute gas-like $\alpha$-condensate. The appropriateness of the THSR approach in describing the Hoyle state is demonstrated by how well the inelastic form factor for transitions between the ground and Hoyle states is reproduced compared with the experimental data. This is a clear indication that the Hoyle state has a large volume, approaching the conditions required for $\alpha$-particle condensation. A complementary way to probe the state's $\alpha$-condensate nature is to show the equivalence between decays to other condensate states; the decay widths for a condensate state should depend entirely on the phase space and Coulomb barrier penetrability for each channel. We have pointed out some flaws in the current approaches and advocate further theoretical work. An upper limit on the direct 3$\alpha$ decay branching ratio of 0.019\% has recently been experimentally measured and will not likely be reduced much further.

In $^{16}$O, the form factor for transitions from the ground state to the 15.1~MeV 0$^+_6$ state has not been measured. This measurement is needed since break-up measurements through the characteristic 4$\alpha$ final state \cite{PhysRevC.100.034320,doi:10.1142/S021773231000068X,Marina} are inconclusive. Beyond oxygen, a high-multiplicity study into the decay of high energy states in $^{28}$Si \cite{PhysRevC.100.034320} assessed the equivalency of all the $\alpha$-condensate decay modes. The results of this work did not provide signatures of $\alpha$-condensate states.

A major unresolved matter is understanding not only the Hoyle state, and Hoyle-like states in heavier systems, but also their excitations. In $^{12}$C, the 0$^+_3$ and 0$^+_4$ have been experimentally measured quite recently \cite{Itoh0+}. A very broad 0$^+$ feature at 10.3 MeV has been known for some time. However, recently Itoh et al. decomposed this into 0$^+_3$ and 0$^+_4$ states at 9.04 and 10.56 MeV. These are interpreted differently; one as $\alpha$-gas state with one $\alpha$-cluster in a higher nodal $s$-state and the other as a linear chain state \cite{Funaki2015}. Furthermore, the structure of the second 2$^+_2$ state \cite{Zimmerman} is still debated. Some consider this as a member of a rotational band built on top the Hoyle state \cite{ACM,DanPRL}. Others speculate this corresponds to a nodal excitation of one of the $\alpha$-particles into a $d$-wave \cite{Funaki2015}.

In an extended THSR approach \cite{THSRBigReview}, where different Gaussian width parameters are permitted, meaning that two of the three $\alpha$-particles can be closer than to the third $\alpha$-particle, a whole spectrum of states in $^{12}$C can be generated. The calculated E2 transition strengths indicate a rotational pattern. However, for 2$^+_2$ $\rightarrow$ 0$^+_2$ and 2$^+_2$ $\rightarrow$ 0$^+_3$, the B(E2) values are similar, meaning that it is not clear whether the 0$^+_2$ or 0$^+_3$ is the band head. The only way to test the validity of this model is to directly measure the $\gamma$ decay of the 2$^+_2$ state to the Hoyle state. Thus far, the only experiment to have unambiguously measured this resonance \cite{Zimmerman} utilised a TPC detector, and the $^{12}$C($\gamma$,$\alpha$) reaction. This reaction was needed to eliminate contributions from nearby broad 0$^+$ resonances. It is possible to contrive an experiment where a $\gamma$ beam of 10~MeV ``on resonance" cleanly populates the 2$^+_2$ state. The $\gamma$ decay could then be inferred by measuring the decay of the Hoyle state into three $\alpha$-particles in the TPC. However, using the B(E2) values published in reference \cite{THSRBigReview} a minuscule branching ratio $\Gamma_\gamma$/$\Gamma_\alpha$ $\approx$ 10$^{-8}$ is calculated, making this measurement impractical at present.

\paragraph{{\bf Funding information}}
This work was funded by the United Kingdom Science and Technology Facilities Council (STFC) under grant number ST/L005751/1 and by the U.S. Department of Energy, Office of Science, Office of Nuclear Physics grants DE-FG02-94ER40870

\end{document}